\documentclass{article}

\usepackage{amssymb,amsmath,latexsym,amsthm,amsfonts}
\usepackage{graphicx}
\usepackage{mathrsfs}
\usepackage{natbib}

\newcommand{\ev}{\mathfrak{Z}}

\newenvironment{theacknowledgments}
     {\section*{Acknowledgements}}
     {\par}

\bibpunct{(}{)}{,}{a}{}{,}

\title{Computational methods for Bayesian model choice}

\author{
{\sc C.P.~Robert}\thanks{CEREMADE - Universit\'e Paris Dauphine, 75775 Paris, and CREST, ENSAE,
France.
Email: \texttt{xian@ceremade.dauphine.fr}}
\and
{\sc D. Wraith}\thanks{CEREMADE - Universit\'e Paris Dauphine, 75775 Paris, France.
Email: \texttt{darren@ceremade.dauphine.fr}}
}

\begin{document}

\maketitle

\begin{abstract}
In this note, we shortly survey some recent approaches on the approximation of the Bayes factor
used in Bayesian hypothesis testing and in Bayesian model choice. In particular, we reassess
importance sampling, harmonic mean sampling, and nested sampling from a unified perspective.

\end{abstract}

\section{Introduction}

The Bayes factor is a fundamental procedure that
stands at the core of the Bayesian theory of testing hypotheses, at least in the approach advocated by \cite{jeffreys:1939}.
(\cite{robert:chopin:rousseau:2009} provides a reassessment of the role of \cite{jeffreys:1939} in setting a
formal framework for Bayesian testing) and by \cite{jaynes:2003}.
Given an hypothesis $H_0:\,\theta\in\Theta_0$ on the parameter $\theta\in\Theta$ of a statistical model,
with density $f(x|\theta)$, under a compatible prior of the form
$$
\pi(\Theta_0) \pi_0(\theta) + \pi(\Theta_0^c) \pi_1(\theta)\,,
$$
the {\em Bayes factor} is defined as the posterior odds to prior odds ratio, namely
\begin{eqnarray*}
B_{01} (x) &=& \displaystyle{ \frac{\pi(\Theta_0|x)}{ \pi(\Theta_0^c|x)} \bigg/
      \frac{\pi(\Theta_0) }{ \pi(\Theta_0^c)} }\\
  &=& { \displaystyle{\int_{\Theta_0} f(x|\theta) \pi_0(\theta) \text{d}\theta} }\bigg/ {
       \displaystyle{\int_{\Theta_0^c} f(x|\theta) \pi_1(\theta) \text{d}\theta} }\,.
\end{eqnarray*}
Since model choice can be considered from a similar perspective, under the Bayesian paradigm (see,
e.g., \cite{robert:2001}), the comparison of models 
$$
\mathfrak{ M}_i : x \sim f_i(x|\theta_i) , \hspace{1cm}  i \in \mathfrak{I}\,,
$$
where the family $\mathfrak{I}$ can be finite or infinite, leads to the same quantities,
$$
{p_i \int_{\Theta_i} f_i(x|\theta_i)
         \pi_i(\theta_i) \text{d}\theta_i}\bigg/
{p_j \int_{\Theta_j} f_j(x|\theta_j) \pi_j(\theta_j)
      \text{d}\theta_j} \,, \quad i,i\in\mathfrak{I} 
$$
In this short survey, we consider some of the most common Monte Carlo solutions used to approximate a generic
Bayes factor or its fundamental component, the {\em evidence}
$${ \ev_k = \int_{\Theta_k} \pi_k(\theta_k) L_k(\theta_k)\,\text{d}\theta_k,}$$
aka the marginal likelihood. Longer entries can be found in \cite{carlin:chib:1995},
\cite{chen:shao:ibrahim:2000}, \cite{robert:casella:2004}, or \cite{friel:pettitt:2008}.
Note that we do not mention here trans-dimensional methods issued from the revolutionary paper
of \cite{green:1995}, since our goal here is to demonstrate that within-model simulation allows for 
the computation of Bayes factors and thus avoids the additional complexity involved in trans-dimensional 
methods.

\section{Importance sampling solutions}

While a regular importance sampling approach is feasible towards the approximation of the Bayes factor
$$
B_{12} = \frac{ \displaystyle{\int_{\Theta_1} f_1(x|\theta_1) \pi_1(\theta_1) \text{d}\theta_1} }{
       \displaystyle{\int_{\Theta_2} f_2(x|\theta_2) \pi_2(\theta_2) \text{d}\theta_2} }\,,
$$
as for instance in
$$
\widehat B_{12} = \frac{ n_1^{-1} \sum_{i=1}^{n_1} f_1(x|\theta_1^i) \pi_1(\theta^i_1)/\varpi_1(\theta^i_1)}
{ n_2^{-1} \sum_{i=1}^{n_2} f_2(x|\theta_2^i) \pi_2(\theta^i_2)/\varpi_2(\theta^i_2)}
$$
which relies on importance functions (densities) $\varpi_1$ and $\varpi_2$ 
and on simulations $\theta_1^i\sim \varpi_1$ and $\theta_2^i\sim\varpi_2$, specific solutions targeted
toward Bayesian model choice are indeed available and preferable. Most of those solutions fit under the
denomination of {\em bridge sampling} and aim at taking advantage of the connections between the two models
under comparison. In fact, when comparing two models with the same complexity (i.e., the same dimension for
their respective parameter spaces), it is often possible to find a reparameterisation of both models in terms
of some specific moments of the sampling model, like $\mathbb{E}[X]$, so that parameters under both models have a common meaning.

\subsection{Bridge sampling}
Assuming that the parameters of both models under comparison, $\theta_1$ and $\theta_2$
respectively, thus belong to the same parameter space (i.e., $\Theta_1=\Theta_2$),
a first solution is to syndicate simulations between both models in order (a) to recycle simulations under one
model for the other model and (b) to create correlation between the estimates of the numerator and of the denominator
of the Bayes factor in order to improve stability in the estimate. This solution is made clear with the formula of \cite{gelman:meng:1998}
$$
B_{12} \approx {1\over n} \sum_{i=1}^n { {\tilde\pi}_1(\theta_{2i}|x) \over
            {\tilde\pi}_2(\theta_{2i}|x) } \,,
$$
when $\theta_{2i} \sim {{\pi_2}}(\theta|x)$, where
\begin{eqnarray*}
\pi_1(\theta_1|x) &\propto& {\tilde\pi}_1(\theta_1|x) \\
\pi_2(\theta_2|x) &\propto& {\tilde\pi}_2(\theta_2|x)\,,
\end{eqnarray*}
as in most Bayesian settings. (An extension to the cases when $\Theta_1\subset\Theta_2$,
including those when the dimension of $\Theta_1$ is smaller than the dimension of $\Theta_2$,
can be easily derived, as shown in \cite{chen:shao:ibrahim:2000}. Note that the assumption 
$\Theta_1=\Theta_2$ signifies that the representations of both models have been reparameterised
in terms of the same moments.)

This is a very special case of the general representation \citep{torrie:valleau:1977}
$$
B_{12} = \dfrac{\mathbb{E}_\varphi\left[\tilde \pi_1(\theta) / \varphi(\theta)\right]}
{\mathbb{E}_\varphi\left[\tilde \pi_2(\theta) / \varphi(\theta)\right]} \,,
$$
which holds for any density $\varphi$ with a sufficiently large support and requires a single
sample $\theta_1,\ldots,\theta_n$ generated from $\varphi$ to produce an importance sampling ratio estimate. 
In that case, a quasi-optimal solution is provided by \cite{chen:shao:ibrahim:2000}, namely $\varphi^*(\theta) 
\propto {\mid \pi_1(\theta)-\pi_2(\theta) \mid}$. The missing normalising constants in both $\pi_1$ and $\pi_2$
obviously mean that this solution cannot be used {\em per se}. In fact, considering the very special case when
$\pi_1(\theta) = \pi_2(\theta)$ on some region of the parameter space, we see that the solution $\varphi^*(\theta)$
should not be used because it is null on some portion of the support of $\pi_1$ and $\pi_2$, thus contradicting a
fundamental requirement of importance sampling.

Another extension of this bridge sampling approach can be based on the general representation
\begin{eqnarray*}
B_{12} &=& { \displaystyle{ \int {\tilde\pi}_2(\theta|x) \alpha(\theta)
   {\pi}_1(\theta|x) d\theta }  \over  \displaystyle{ \int {\tilde\pi}_1
   (\theta|x) \alpha(\theta) {\pi}_2(\theta|x) d\theta } }
           \\
  &\approx& { \displaystyle{ {1\over n_1} \sum_{i=1}^{n_1}
  {\tilde\pi}_2(\theta_{1i}|x) \alpha(\theta_{1i}) }  \over
  \displaystyle{ {1\over n_2} \sum_{i=1}^{n_2} {\tilde\pi}_1(\theta_{2i}|x)
  \alpha(\theta_{2i}) } }
\end{eqnarray*}
where $\theta_{ji} \sim \pi_j(\theta|x)$, which applies for any positive and integrable function $\alpha$.
Some choices of $\alpha$ do lead to very poor performances of the method in connection with the
harmonic mean approach \citep{raftery:etal:2008}, but there
exists a quasi-optimal solution, as provided by \cite{gelman:meng:1998}:
$$
{\alpha^\star \propto \dfrac{1}{n_1{\pi}_1(\theta|x) +n_2  {\pi}_2(\theta|x)}} \,.
$$
Once again, the optimum cannot be used {\em per se}, since it requires the normalising constants of
both $\pi_1$ and $\pi_2$. As suggested by \cite{gelman:meng:1998},
an approximate version uses iterative versions of $\alpha^\star$, based on successive iterates
of approximations to the Bayes factor. Note that this solution recycles simulations from each posterior, which is
quite appropriate since one model is selected via the Bayes factor, instead of using an importance sample common to
both approximations. We will see below an alternative representation of the bridge factor that bypasses this difficulty.

\subsection{Harmonic means}
\begin{figure}
\centerline{\includegraphics[width=.5\textwidth]{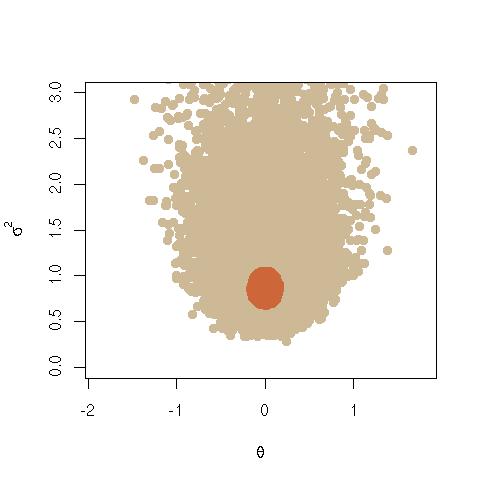}\includegraphics[width=.5\textwidth]{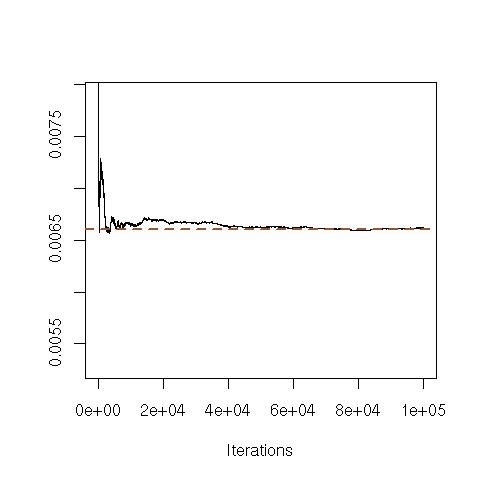}}
\caption{\label{fig:hpd}
{\em (left)} Representation of a Gibbs sample of $10^3$ parameters $(\theta,\sigma^2)$ for the normal model, 
$x_1,\ldots,x_n\sim\mathcal{N}(\theta,\sigma^2)$ with 
$\overline x=0$, $s^2=1$ and $n=10$, under Jeffreys' prior,
along with the pointwise approximation to the $10\% $ HPD region {\em (in darker hues)}.
{\em (right)} Evaluation of the approximation of the evidence based on the density $\varphi$ for the uniform
distribution on the ellipse approximating this HPD region.}
\end{figure}

While using the generic harmonic mean approximation to the marginal likelihood is often fraught with danger
\citep{neal:1994,chopin:robert:2007c}, the representation \citep{gelfand:dey:1994}
\begin{equation}\label{eq:harmony}
\mathbb{E}^{\pi_k}\left[\left.\frac{\varphi(\theta_k) }{\pi_k(\theta_k)L_k(\theta_k)}\right| x \right]
= \int \frac{\varphi(\theta_k) }{\pi_k(\theta_k)L_k(\theta_k)} \,
\frac{\pi_k(\theta_k)L_k(\theta_k)}{\mathfrak{Z}_k}\,\text{d}\theta_k
= \frac{1}{\mathfrak{Z}_k}
\end{equation}
holds, no matter what the density $\varphi(\cdot)$ is. This representation is remarkable in that it allows
for a direct processing of Monte Carlo or MCMC output from the posterior distribution. In addition, and as 
opposed to usual importance sampling constraints, the density
$\varphi(\theta)$ must have lighter---rather than fatter---tails than $\pi_k(\theta_k)L_k(\theta_k)$ 
for the approximation of the Bayes factor
$$
1\Bigg/ \frac{1}{T}\,\sum_{t=1}^T \frac{\varphi(\theta_k^{(t)})
        }{\pi_k(\theta_k^{(t)})L_k(\theta_k^{(t)})}
$$
to enjoy finite variance. Therefore, using $\varphi(\theta_k)=\pi_k(\theta_k)$ as in the original harmonic mean approximation
\citep{newton:raftery:1994} will often result in an infinite variance, as discussed by \citep{neal:1994}. 
On the opposite, using $\varphi$'s with constrained supports derived from a Monte Carlo sample,
like the convex hull of the simulations corresponding to the $10\%$ or to the $25\%$ HPD regions---that again is easily derived
from the simulations---is both completely appropriate and implementable, as illustrated by Figure \ref{fig:hpd} for a toy example.
In this example, we used the simulations within the HPD region to define an ellipse and consequently a uniform density $\varphi$
over this ellipse. Since the true ``evidence" can be computed analytically, checking the convergence of the harmonic approximation
is straightforward. (We warn the reader that this ``evidence" cannot be used in a model comparison framework, because it is associated
with an improper prior that is not acceptable in testing settings. See \cite{degroot:1973} or \cite{robert:2001} for more details.
Nonetheless, it provides a valid toy example to check the convergence of an integral approximation.)

\subsection{Mixture bridge sampling}
\begin{figure}
\centerline{\includegraphics[width=.5\textwidth]{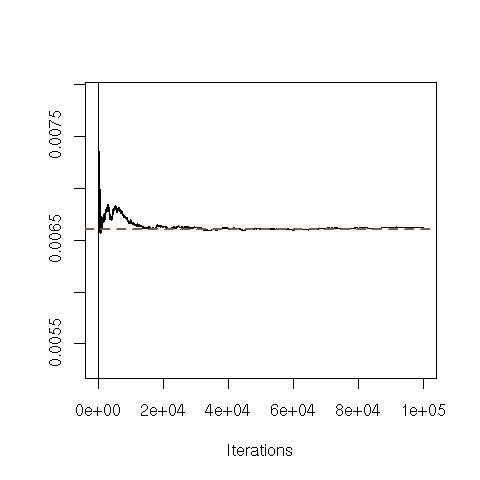}\includegraphics[width=.5\textwidth]{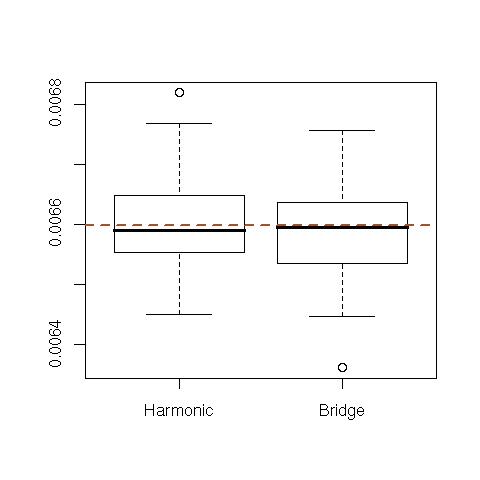}}
\caption{\label{fig:hpdbis}
For the same setting as Figure \ref{fig:hpd},
{\em (left)} Evaluation of the bridge sampling approximation of the evidence based on the mixture
$\tilde\varphi$ when the density $\varphi$ corresponds to the uniform
distribution on the ellipse approximating the same HPD region as in Figure \ref{fig:hpd}, using a value
of $\omega$ equal to one-tenth of the true evidence;
{\em (right)} boxplot comparison of the variations of both approaches based on $100$ Monte Carlo replicas 
with samples of size $10^4$. On both graphs, the true evidence is represented by the horizontal dotted line.}
\end{figure}

As noted above, a remarkable feature of the representation \eqref{eq:harmony} is that the derived implementation
can directly exploit the output of any MCMC sampler. Another approach introduced in \cite{chopin:robert:2007} aims
at the same goal and attains the optimal bridge sampler from a completely different perspective. It considers a 
specific mixture structure as importance function, of the form
$$
\widetilde \varphi(\theta) \propto \omega \pi(\theta)L(\theta) + \varphi(\theta)\,,
$$
where $\varphi(\cdot)$ is an arbitrary but fully normalised density. Simulating from this mixture, assuming there already
exists an MCMC sampler with stationary distribution $\pi(\theta|x)\propto \pi(\theta)L(\theta)$, is straightforward, thanks 
to a tailored Gibbs sampler:

\medskip
\noindent{\bf Mixture Gibbs sampler}\\
{\sffamily
At iteration $t$
\begin{enumerate}
\item Take $\delta^{(t)}=1$ with probability
$$
\omega_1 \pi_k(\theta_k^{(t-1)})L_k(\theta_k^{(t-1)})\bigg/
\left(\omega_1 \pi_k(\theta_k^{(t-1)})L_k(\theta_k^{(t-1)})
+ \varphi(\theta_k^{(t-1)})\right)
$$
and $\delta^{(t)}=2$ otherwise;
\item If $\delta^{(t)}=1$, generate $\theta_k^{(t)}\sim\text{MCMC}(\theta_k^{(t-1)},\theta_k)$
where $\text{MCMC}(\theta_k,\theta_k^\prime)$ denotes an arbitrary MCMC kernel associated with the
posterior $\pi_k(\theta_k|x)\propto \pi_k(\theta_k)L_k(\theta_k)$;
\item If $\delta^{(t)}=2$, generate $\theta_k^{(t)}\sim\varphi(\theta_k)$ independently
\end{enumerate}
}
The simulation step 1. selecting between both components of the mixture is not only allowing for simulation
from this mixture, but it also does provide a direct estimate to the evidence. Indeed, the Rao-Blackwellised estimate
$$
\hat{\xi}=\frac{1}{T}\,\sum_{t=1}^T  \omega_1 \pi_k(\theta_k^{(t)})L_k(\theta_k^{(t)}) \bigg/  \omega_1
\pi_k(\theta_k^{(t)})L_k(\theta_k^{(t)}) + \varphi(\theta_k^{(t)}) \,,
$$
converges to $\omega_1 \mathfrak{Z}_k/\{ \omega_1 \mathfrak{Z}_k + 1\}$
and we can thus deduce  
$$
\hat{\ev}_{3k} = \frac{1}{\omega_1}\,\frac{\sum_{t=1}^T \omega_1 \pi_k(\theta_k^{(t)})L_k(\theta_k^{(t)}) \bigg/  \omega_1
\pi(\theta_k^{(t)})L_k(\theta_k^{(t)}) + \varphi(\theta_k^{(t)})}{
\sum_{t=1}^T \varphi(\theta_k^{(t)}) \bigg/  \omega_1
\pi_k(\theta_k^{(t)})L_k(\theta_k^{(t)}) + \varphi(\theta_k^{(t)})}\,.
$$
We have thus recovered the optimal bridge sampling estimate from this different perspective, akin to
\cite{bartscami}, that is more in line with reversible jump trans-dimensional schemes than regular
importance sampling. The only modification compared with the original version is that the sequence
$(\theta_k^{(t)})$ is generated from the mixture for both the numerator and the denominator. Figure
\ref{fig:hpdbis} shows that, for the toy example introduced in Figure \ref{fig:hpd}, the harmonic mean
approximation does as well as the optimal bridge sampling solution.

\section{Nested sampling}
This method introduced in \cite{skilling:2007a,skilling:2007b} (although an earlier version can be
found in \cite{burrows:1980}) produces a very specific type of importance sampling based on 
constrained simulations from the prior distribution. While more details descriptions are available in
the above reference and \cite{chopin:robert:2007} (as well as a full convergence assessment in the later paper), let us
recall here that nested sampling is based on the one-dimensional representation
$$ \ev = \mathbb{E}^\pi[L(\theta)] = \int_0^1 \varphi(x)\,\text{d}x$$
of the evidence, when
$$ \varphi^{-1}(l) = P^\pi(L(\theta)>l)$$
is the survival probability function associated with the likelihood.
The approximation of $\ev$ by a Riemann sum:
$$\widehat\ev = \sum_{i=1}^N (x_{i-1}-x_i)\varphi(x_i)$$
where the $x_i$'s are either deterministic, e.g.~$x_i=\exp\{-i/N\}$, or random, 
also allows for the representation 
$$
\widehat\ev = \sum_{i=0}^{N-1} \{\varphi(x_{i+1})-\varphi(x_i)\}x_i
$$
which is a special case of
\begin{equation}\label{eq:otherside}
\widehat\ev = \sum_{i=0}^{N-1} \{L(\theta_{(i+1)})-L(\theta_{(i)})\}\pi(\{\theta;L(\theta)>L(\theta_{(i)})\})
\end{equation}
where $\cdots L(\theta_{(i+1)}) > L(\theta_{(i)})\cdots$.
(This can be seen as the Lebesgue version of the Riemann's sum, the triangulation being on the
second axis instead of the first axis.)
Since $\varphi$ is rarely available in closed form, the nested sampling algorithm relies on an estimate of $\varphi(x_i)$
or, equivalently, of $\pi(\{\theta;L(\theta)>L(\theta_{(i)})\})$:

\medskip
\noindent{\bf Nested sampling algorithm}\\
{\sffamily
Start with $N$ values $\theta_{1},\ldots,\theta_{N}$ sampled from $\pi$\\
At iteration $i$,
\begin{enumerate}
 \item Take $\varphi_i=L(\theta_{k})$, where $\theta_k$ is the point with smallest likelihood in
the pool of $\theta_i$'s
 \item Replace $\theta_{k}$ with a sample from the prior {constrained to $L(\theta)>\varphi_i$}:
the current $N$ points are sampled from {prior constrained to $L(\theta)>\varphi_i$.}
\end{enumerate}
}

In terms of the representation \eqref{eq:otherside}, this amounts to use the approximation
$$
\widehat\pi(\{\theta;L(\theta)>L(\theta_{(i)})\})/\pi(\{\theta;L(\theta)>L(\theta_{(i-1)})\}) = (N-1) / N\,.
$$
As discussed in \cite{evans:2007} and \cite{chopin:robert:2007b}, the dominating term in the approximation
is the stochastic part that converges at $\text{O}(\sqrt{n})$ speed. The method thus formally compares with
those mentioned in the previous section. At a more practical level, nested sampling can be interpreted as
an importance sampling technique where, instead of simulating a whole sample from the prior distribution,
$\pi(\theta)$, and approximating the evidence by
$$
\frac{1}{N}\,\sum_{i=1}^N L(\theta_i)\,,
$$
which is usually inefficient,
points $\theta_{(i)}$ are successively sampled from the prior restricted to higher and higher levels of the likelihood, with 
decreasing weights 
$$
\frac{1}{N}\,\left(1-\frac{1}{N}\right)^{i-1}\,.
$$

\begin{figure}
\centerline{\includegraphics[width=.4\textwidth]{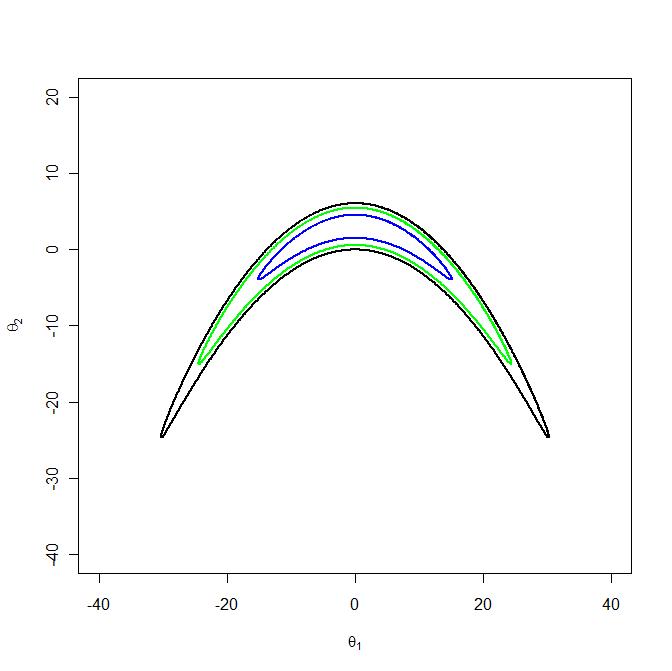}}
\caption{\label{fig:banana}
Contours (at the levels $68\%$, $95\%$ and $99.9\%$) of the likelihood 
of the twisted normal model for $\beta=.03,\ \sigma_1^2=(100)$.}
\end{figure}

To illustrate the comparison with a standard importance sampling approximation,
we now consider an artificial example based on the likelihood of a twisted normal distribution 
(first introduced in \citep{haario:sacksman:tamminen:1999} as a benchmark for adaptive MCMC
schemes) in two dimensions with covariance matrix $\Sigma=\textup{diag}(\sigma_1^2,1)$.  
The ``twist" is due to considering the transform $\theta_2^\prime=\theta_2+\beta(\theta_1^2-\sigma_1^2)$, 
which leads to a sharp bend in the likelihood contours, as shown in Figure \ref{fig:banana}. Since the Jacobian of the twist is equal to~1, the density is thus defined as:
$$
  \psi(\theta_1,\theta_2)=(\theta_1,\theta_2+\beta(\theta_1^2-\sigma_1^2)) \sim {\cal N}_2(0,\Sigma) \,.
$$
If we consider $\beta$ and $\sigma_1^2$ as known, the appeal of this example is in integrating 
over the parameters $\theta_1$ and $\theta_2$ with priors $\pi(\theta_1,\theta_2)$. The toy evidence can thus be represented as
$$
  \ev_1=\int \psi(\theta_1,\theta_2)\pi(\theta_1,\theta_2) \text{d}\theta
$$
For the example that we consider, we fix $\beta=0.03$,  $\sigma_1^2=100$ (as represented in Figure~\ref{fig:banana}) and 
we use flat priors on $\theta_1$ in $(-40,40)$ and on $\theta_2$ also on $(-40,40)$. 
(The prior square is chosen arbitrarily to allow all possible values and still to retain a compact parameter space. Furthermore, a flat prior allows for 
an easy implementation of nested sampling since the constrained simulation can be implemented via a random walk move, as pointed out in 
\cite{skilling:2007a}.).  Integrating the likelihood over this region of the parameter space presents a challenging problem for any approach as 
the coverage of the tails of the twisted normal distribution can be difficult or even impossible to capture. (This point is discussed at length in
\cite{wraith:etal:2009}.)

In this toy example, the two-dimensional nature of the parameter space does allow for a numerical integration of $\ev_1$, thus producing
a reference value, based on a Riemann approximation of the integral and a grid of 1000$\times$1000 points in the (-40, -40)$\times$(40,40)
square (an adaptive quadrature approach was also used as a check). This approach leads to a stable evaluation of $\ev_1$ that can be 
reliably taken as the reference against which we test alternative approximation methods.

The comparison is here restricted to a standard nested sampling algorithm derived from \cite{skilling:2007a} 
and a population Monte Carlo (PMC) mixture importance sampler constructed in \cite{wraith:etal:2009} for this
benchmark problem and introduced in \cite{cappe:douc:guillin:marin:robert:2007} in a general framework. 
Briefly, this importance sampling approach is adaptive as it
consists of modifying the parameters of the importance function (a mixture density), bringing it closer to the posterior density 
over a small number of iterations. The proximity is measured in terms of the Kullback divergence between the posterior density and importance function,
since utilising an integrated EM approach ensures that the divergence successively decreases at each iteration. While this adaptive importance 
sampler derives a proposal $\varphi$ to simulate from the (pseudo-)posterior $\psi(\theta_1,\theta_2)\pi(\theta_1,\theta_2)$, it can obviously 
provide in addition an approximation of the marginal likelihood $\ev_1$. (We stress that any importance sampler used in this setting offers this
facility of providing an approximation of both the evidence and of its variability.)

For comparison purposes, the PMC approach uses 5000 simulated points per iteration over a total of 10 iterations and then a final
sample of 50000 points, simulated from the ``optimal" importance sampling function obtained through the PMC sequence. 
The importance function to be optimised consists of a mixture of 9 multivariate Student t's with 9 degrees of freedom for each component.  
For the initial values of the importance function, components of the mixture are located randomly in different directions slightly away from 0: 
the mean of the components are drawn from a bivariate Gaussian distribution with mean $0$ and covariance equal to $\Sigma_0/5$,
where $\Sigma_0$ is a diagonal matrix with diagonal entries $(200, 50)$.
In parallel, we run the nested sampling algorithm with N = 1000 initial points, reproducing the implementation of \cite{skilling:2007a},
using 50 steps of a random walk in ($\theta_1,\theta_2$) constrained by the likelihood
boundary to produce the next value, based on the contribution of the current value of ($\theta_1,\theta_2$) to the approximation
of $\ev_1$. The step size (ie.~the variance) in the random walk is 0.1 and the process is repeated (iterated) 10,000 times (and monitored) to ensure 
a definitive completion of the accumulation of $\ev_1$. Alternative scenarios, including changes to the number of points $N$, 
to the number of steps and to the step size were explored to assess the sensitivity of the results to the values set, but 
they did not lead to an improvement in the results.  To assess the variability of the results, 100 simulation runs for both PMC and nested sampling 
algorithms have been used.

\begin{figure}
\centerline{\includegraphics[width=.35\textwidth]{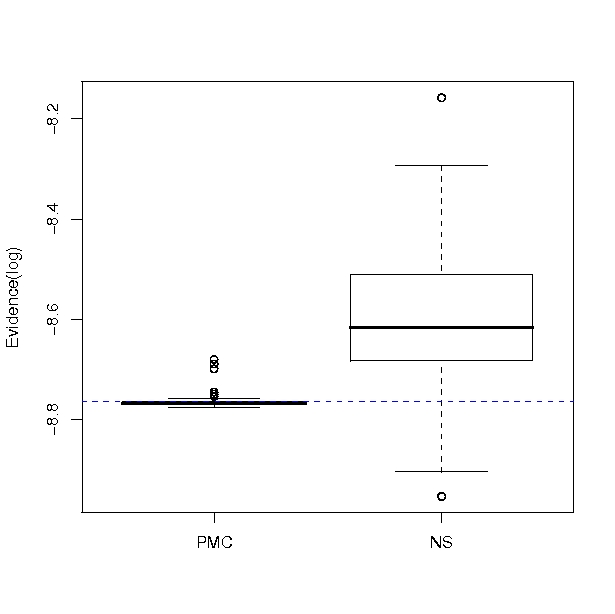}
\includegraphics[width=.35\textwidth]{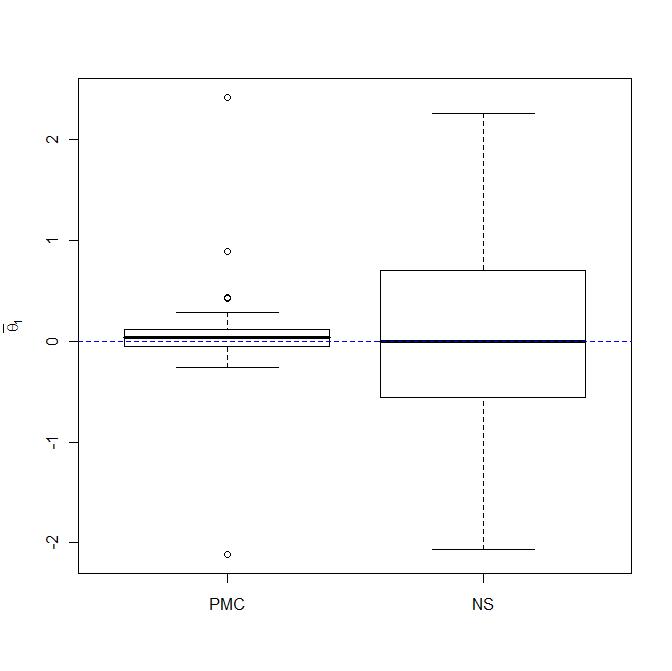} \includegraphics[width=.35\textwidth]{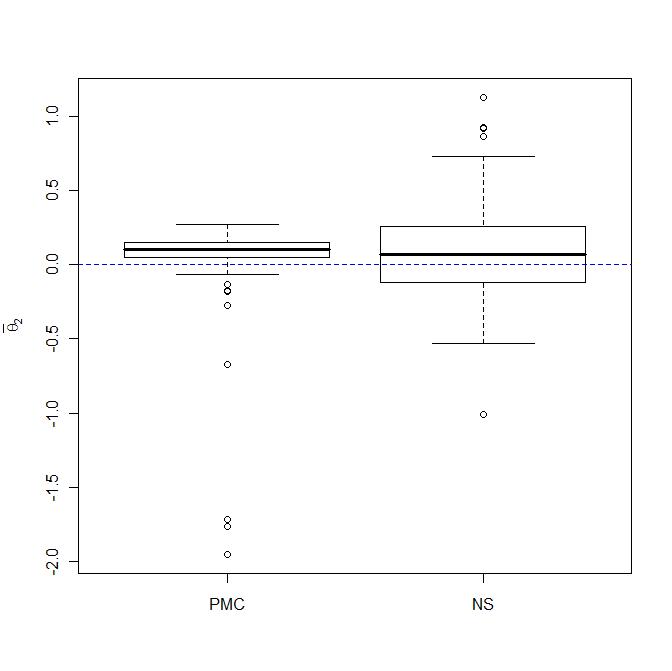}}
\caption{\label{fig:apple}
Comparison of nested sampling with PMC over 100 simulation runs for
{\em (left)} Evidence estimation;
{\em (centre)} $\mathbb{E}[\theta_1]$ estimation;
{\em (right)} $\mathbb{E}[\theta_2]$ estimation.
The true value is represented as an horizontal dotted line.}
\end{figure}

Figure~\ref{fig:apple} summarises our results for PMC compared with nested sampling over the 100 simulation runs for evaluation of the evidence $\ev_1$ and 
of the posterior mean for $\theta_1$ ($\mathbb{E}[\theta_1]$) and $\theta_2$ ($\mathbb{E}[\theta_1]$), since the outcome of a nested sampling run can be
utilised as any importance sampling output. Those results suggest that nested sampling exhibits a slight upward bias for the evaluation of the evidence 
(a point also noted in \cite{chopin:robert:2007}) while it approximately produces the same numerical value for 
the estimates of $\mathbb{E}[\theta_1]$ and of $\mathbb{E}[\theta_2]$, albeit with a greater variability.

\section{Comments}
Various importance sampling strategies have been proposed recently that explicitly target the evidence. While there is no clear winner emerging from the
comparison, we conclude that the bridge sampling strategy remains a reference in this domain, but also that the harmonic version of \cite{gelfand:dey:1994,
bartscami,chopin:robert:2007} may produce valuable approximations if the empirical HPD regions are exploited in the way described in the current paper.


\begin{theacknowledgments}
Both C.P.~Robert and D.~Wraith are supported by the 2006-2009 ANR ``Ecosstat". C.P.~Robert is grateful to the
organisers of MaxEnt 2009 for their kind invitation and to O.~Capp\'e for discussions, as well as to
the students attending the course on Bayesian Data Analysis for Ecologists
in Gran Paradiso National Park, Aosta, Italy, for they made him realise the strong potential of
using empirical HPD regions.
\end{theacknowledgments}



\bibliographystyle{aipproc}   

\bibliography{biblio}

\IfFileExists{\jobname.bbl}{}
 {\typeout{}
  \typeout{******************************************}
  \typeout{** Please run "bibtex \jobname" to obtain}
  \typeout{** the bibliography and then re-run LaTeX}
  \typeout{** twice to fix the references!}
  \typeout{******************************************}
  \typeout{}
 }

\end{document}